\documentclass[
    ,final            
  ]
  {aipproc}

\usepackage{epsfig}
\layoutstyle{6x9}

\begin{document}

\title{Exploring Nucleon Spin Structure in Longitudinally Polarized Collisions}

\author{Marco Stratmann}{address={Inst.\ for Theor.\ Physics, Univ.\ of Regensburg, 
D-93040 Regensburg, Germany\\ E-mail: marco.stratmann@physik.uni-regensburg.de}}

\begin{abstract}
We review how RHIC is expected to deepen our understanding of the 
spin structure of longitudinally polarized nucleons.
After briefly outlining the current status of spin-dependent
parton densities and pointing out open questions, we focus on theoretical calculations 
and predictions relevant for the RHIC spin program.
Estimates of the expected statistical accuracy for such measurements are presented,
taking into account the acceptance of the RHIC detectors.
\end{abstract}

\maketitle

\section{Lessons from polarized deep inelastic scattering}
%
Before reviewing the prospects for spin physics at the BNL-RHIC we briefly
turn to longitudinally polarized deep-inelastic scattering (DIS)
and what we have learned from more than 
twenty years of beautiful data~\cite{ref:expdata}.
To next-to-leading order (NLO) in the strong coupling $\alpha_s$, 
the DIS structure function $g_1$, which parametrizes our ignorance about the nucleon
spin structure, can be expressed as
\begin{equation}
g_1(x,Q^2) = \frac{1}{2} \sum_{q=u,d,s} e_q^2 \left[ \left(\Delta q + \Delta \bar{q} \right)
\otimes \left( 1+\frac{\alpha_s}{2\pi} \Delta C_q \right) + 
\frac{\alpha_s}{2\pi} \Delta g \otimes \Delta C_g \right] (x,Q^2)\;\;\;.
\end{equation}
The symbol $\otimes$ denotes the usual convolution, and $\Delta C_{q,g}$ are the perturbatively
calculable coefficient functions, which are known even up to next-to-next-to-leading order
\cite{ref:neerven}. The $\Delta f$, $f= (q,\, \bar{q},\, g)$, 
are the spin-dependent parton distributions, defined as
\begin{equation}
\label{eq:pdfs}
\Delta f(x,\mu) \equiv f^+(x,\mu) - f^-(x,\mu)\;\;\;,
\end{equation}
where $f^+$ $(f^-)$ is the number density of a parton type $f$ with helicity ``$+$'' (``$-$'')
in a proton with positive helicity, carrying a fraction $x$ of the proton's momentum.
Once they are known at some initial scale $\mu_0$, their
scale $\mu$ dependence is governed by a set of evolution equations
\begin{equation}
\label{eq:evol}
\mu\frac{{d}}{{d}\mu} \left( \!\! \begin{array}{c} {\Delta q(x,\mu)} \\
                                              {\Delta g(x,\mu)} \end{array} \!\! \right)
= \left(\!\! \begin{array}{cc}
 \Delta{\cal P}_{qq} &  \Delta{\cal P}_{qg} \\
 \Delta{\cal P}_{gq} &  \Delta{\cal P}_{gg}
                                \end{array}  \!\! \right) \otimes
\left( \!\! \begin{array}{c} {\Delta q} \\ {\Delta g} \end{array} \!\! \right)
(x,\mu )\;\;\;.
\end{equation}
So far, the spin-dependent $j\to i$ splitting functions entering these evolution equations
have been calculated up to NLO accuracy \cite{ref:nlokernel}.
%
%

\begin{figure}[th]
\vspace*{-0.5cm}
\centerline{\epsfxsize=0.59\textwidth\epsfbox{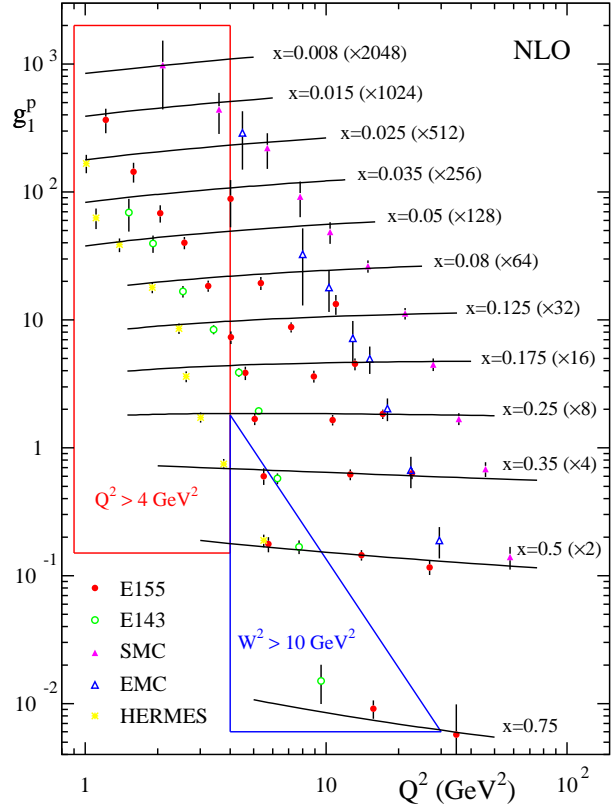}}   
\caption{Available information on $g_1(x,Q^2)$ as collected by 
fixed-target experiments \cite{ref:expdata} compared to results of a 
typical NLO QCD fit (solid lines).
The indicated rectangular and triangular regions contain data
which would not pass kinematical cuts of $Q^2>4\,\mathrm{GeV}^2$ 
and $W^2>10\,\mathrm{GeV}^2$, respectively, usually imposed in all fits
to unpolarized DIS data. \label{fig:fig1}}
\end{figure}
Figure~\ref{fig:fig1} compares the available information on $g_1(x,Q^2)$ 
for DIS off a proton target to results of a typical NLO QCD fit.
From such types of analyses a pretty good knowledge of certain 
combinations of different quark flavors has emerged, and it became clear 
that quarks contribute only a small fraction to the proton's spin.
However, there is still considerable lack of knowledge regarding the
polarized gluon density $\Delta g$, which is basically unconstrained
by present data, the separation of quark and antiquark densities and of different
flavors, and the orbital angular momentum of quarks and gluons inside a nucleon.
With the exception of orbital angular momentum RHIC can address all of these questions as will be
demonstrated in the following~\cite{ref:rhicreport}.

There is also an important difficulty when analyzing polarized DIS data in terms
of spin-dependent parton densities: compared to the unpolarized case
the presently available kinematical coverage in $x$ and $Q^2$ and the statistical
precision of polarized DIS data are much more limited~\cite{ref:expdata}. 
As a consequence, one is forced to include data into the fits from $(x,Q^2)$-regions where
corresponding fits of unpolarized leading-twist parton densities 
start to break down, see Fig.~\ref{fig:fig1}.
Data from RHIC, taken at ``resolution'' scales $\mu$ where perturbative QCD and 
the leading-twist approximation are supposed to work, can shed light on the 
possible size of unwanted higher-twist contributions in presently available sets of 
polarized parton distributions. 

\section{Spin Physics at RHIC}
%
\subsection{Prerequisites}
\begin{figure}[ht]
\vspace*{-0.2cm}
\epsfig{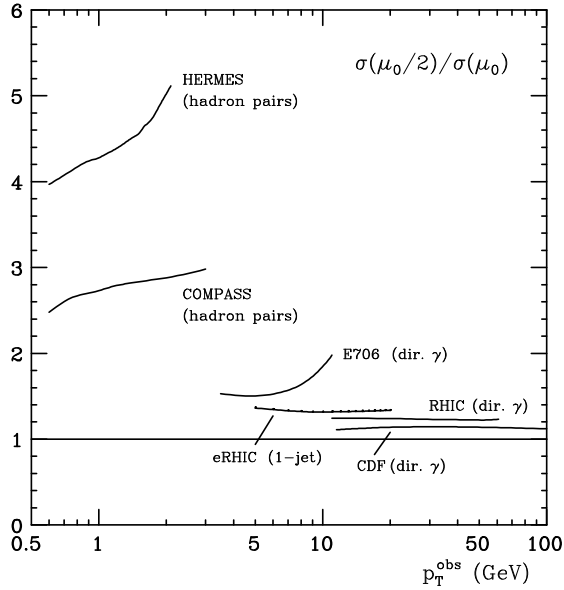}
\vspace*{-0.2cm}
\caption{Typical factorization scale dependence for various ``high-$p_T$'' processes 
and experiments as a function of $p_T$.
Shown is the cross section ratio for two choices of scale, $p_T$ and $p_T/2$. \label{fig:fig2}}
\vspace*{-0.4cm}
\end{figure}

%
The QCD-improved parton model has been successfully applied to many high energy 
scattering processes. The predictive power of perturbative QCD follows from the 
universality of the parton distribution and fragmentation functions which is based on
the factorization theorem. To be specific, let us consider the inclusive production of 
a hadron $H$, e.g., a pion, in longitudinally polarized $pp$ collisions at a c.m.s. energy
$\sqrt{S}$.
The cross section can be written in a factorized form as a convolution of perturbatively 
calculable partonic cross sections $d\Delta\hat{\sigma}_{ab}^c$ describing the hard scattering
$a b \to c X$ and appropriate combinations of parton densities $\Delta f_{a,b}$ and 
fragmentation functions $D_c^H$ embodying the non-perturbative physics:
\begin{equation}
\label{eq:eq1}
\frac{d\Delta\sigma^H}{d\Gamma} = \sum_{abc} \int dx_a\, dx_b\,dz\, \Delta f_a(x_a,\mu)\,
\Delta f_b(x_b,\mu)\,\frac{d\Delta\hat{\sigma}_{ab}^c}{d\Gamma}(x_a,x_b,z,S,{\Gamma},\mu)\,
D_c^H(z,\mu)\;.
\end{equation}
Here, $\Gamma$ stands for any appropriate set of kinematical variables like
the transverse momentum $p_T$ and/or rapidity $y$ of the observed hadron. 
The $D_c^H$ are the parton-to-hadron fragmentation functions. Their scale $\mu$-dependence 
is governed by a set of equations very similar to (\ref{eq:evol}).
The factorization scale $\mu$, introduced on the r.h.s.\ of (\ref{eq:eq1}), separates 
long- and short-distance phenomena. $\mu$ is completely arbitrary but usually chosen to be 
of the order of the scale characterizing the hard interaction, for instance $p_T$. 
Since the l.h.s.\ of (\ref{eq:eq1}) has to be independent of $\mu$
(and other theoretical conventions), any residual
dependence of the r.h.s.\ on the actual choice of $\mu$ gives an indication of how well
the theoretical calculation is under control and can be trusted.
In particular, leading order (LO) estimates suffer from a strong, uncontrollable 
scale dependence and hence are not sufficient for comparing theory with data.
Figure~\ref{fig:fig2} shows a typical factorization scale dependence for 
various ``high-$p_T$'' processes and experiments as a function of $p_T$. Clearly, the situation
is only acceptable at collider experiments where one can easily reach $p_T$
values in excess of 5 GeV. $p_T$ values of the order of 1-2~GeV, accessible 
at fixed-target experiments, are not sufficient to provide a large enough scale $\mu$ for
performing perturbative calculations reliably. 
For simplicity we have not distinguished between 
renormalization and initial/final-state factorization scales 
in (\ref{eq:eq1}) which can be chosen differently.

In practice, spin experiments do not measure the polarized cross section $d\Delta \sigma/d\Gamma$
itself, but the longitudinal spin asymmetry, which is given by the ratio of the 
polarized and unpolarized cross sections, i.e., for our example
above, Eq.~(\ref{eq:eq1}), it reads
\begin{equation}
\label{eq:eq2}
A_{\mathrm{LL}}^H\equiv \frac{d\Delta \sigma^H/d\Gamma}{d\sigma^H/d\Gamma}\;\;\;.
\end{equation}
The unpolarized cross section $d\sigma^H/d\Gamma$ is given by Eq.~(\ref{eq:eq1})
with all polarized quantities replaced by their unpolarized counterparts.
At RHIC one can also study doubly transverse spin asymmetries \cite{ref:rhicreport} but
here we will focus on longitudinal polarization only.

\subsection{Accessing $\Delta g$}
The main thrust of the RHIC spin program \cite{ref:rhicreport} is to pin down the so far elusive
gluon helicity distributions $\Delta g(x,\mu)$. The strength of RHIC is the possibility
to probe $\Delta g(x,\mu)$ in a variety of hard processes \cite{ref:rhicreport},
in each case at sufficiently large $p_T$ where perturbative QCD is expected to work. 
This not only allows to determine the $x$-shape of $\Delta g(x,\mu)$ for 
$x\;$\raisebox{-1.5mm}{${\stackrel{\textstyle >}{\sim}}$}$ 0.01$ but 
also verifies the universality property of polarized parton 
densities for the first time. In the following we review the status of theoretical
calculations for processes sensitive to $\Delta g$, experimental aspects can be 
found, e.g., in \cite{ref:bland}.

The ``classical'' tool for determining the gluon density is high-$p_T$ prompt photon
production due to the dominance of the LO Compton process, $qg \to \gamma q$.
Exploiting this feature, both RHIC experiments, PHENIX and STAR, intend to use this process
for a measurement of $\Delta g$. 
Apart from ``direct'' mechanisms like $qg \to \gamma q$, the photon can also be
produced by a parton, scattered or created in a hard QCD reaction, which fragments
into the photon. Such a contribution naturally arises in a QCD calculation from the
necessity of factorizing final-state collinear singularities into a photon
fragmentation function $D_c^\gamma$. However, since photons produced through fragmentation
are always accompanied by hadronic debris, an ``isolation cut'' imposed on the photon signal
in experiment, e.g., a ``cone'', strongly reduces such contributions to the cross section.

\begin{figure}[th]
\vspace*{-0.3cm}
\centerline{\epsfxsize=3.15in\epsfbox{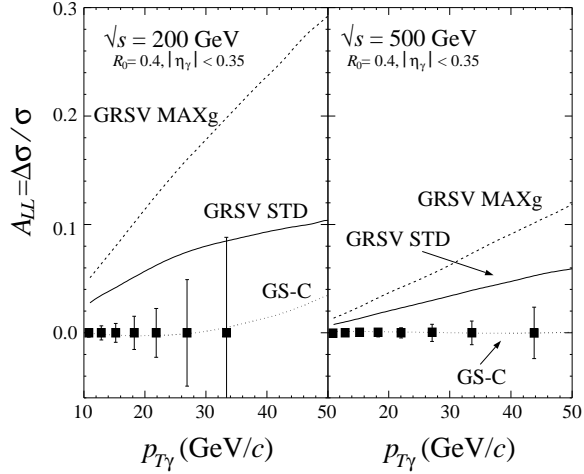}}   
\caption{$A_{\mathrm{LL}}$ for prompt photon production in NLO QCD as a
function of $p_T$ for different sets of parton densities.
The ``error bars'' indicate the expected statistical accuracy
$\delta A_{\mathrm{LL}}$, Eq.~(\ref{eq:error}),
for the PHENIX experiment. Figure taken from \cite{ref:frixione}. \label{fig:fig3}}
\end{figure}
The NLO QCD corrections to the direct (non-fragmentation) processes have been
calculated in \cite{ref:nlophoton} and lead to a much 
reduced factorization scale dependence as compared
to LO estimates. In addition, Monte Carlo codes have been 
developed \cite{ref:mcphoton,ref:frixione}, 
which allow to include various isolation criteria and to study also photon-plus-jet
observables. The latter are relevant for $\Delta g$ measurements planned at 
STAR \cite{ref:rhicreport,ref:bland}.
Since present comparisons between experiment and theory are not fully satisfactory in the
unpolarized case, in particular in the fixed-target regime, considerable efforts have been 
made to push calculations beyond the NLO of QCD by including resummations of large
logarithms \cite{ref:resum}. It is hence not unlikely that a better understanding of
prompt photon production can be achieved soon.
Figure~\ref{fig:fig3} shows $A_{\mathrm{LL}}^\gamma$ as predicted by a NLO QCD
calculation \cite{ref:frixione} as a function of the photon's transverse momentum
$p_T$. The applied rapidity cut $|\eta|\leq 0.35$ matches the acceptance of the PHENIX detector.
The important result is that the expected statistical errors $\delta A_{\mathrm{LL}}$
are considerably smaller than
the changes in $A_{\mathrm{LL}}^\gamma$ due to different spin-dependent gluon 
densities over a wide range of $p_T$. Hence RHIC should be able to 
probe $\Delta g$ in prompt photon production.
$\delta A_{\mathrm{LL}}$ may be estimated by the formula
\begin{equation}
\label{eq:error}
\delta A_{\mathrm{LL}} = \frac{1}{P^2 \sqrt{{\cal{L}}\sigma_{\mathrm{bin}}}}\;\;,
\end{equation}   
where $P$ is the polarization of one beam, ${\cal{L}}$ the integrated luminosity of the
$pp$ collisions, and $\sigma_{\mathrm{bin}}$ the unpolarized cross section integrated over
the $p_T$-bin for which the error is to be determined.
Unless stated otherwise, $P=0.7$ and ${\cal{L}}=320\, (800)\;\mathrm{pb}^{-1}$ is used
in Eq.~(\ref{eq:error}) for 
$pp$ collisions at $\sqrt{S}=200\, (500)\;\mathrm{GeV}$ \cite{ref:rhicreport}.

Jets are another key-process to pin down $\Delta g$ at RHIC: they are copiously
produced at $\sqrt{S}=500\,\mathrm{GeV}$, even at high $p_T$,
$15\; $\raisebox{-1.5mm}{${\stackrel{\textstyle <}{\sim}}$}$ \; p_T \;
$\raisebox{-1.5mm}{${\stackrel{\textstyle <}{\sim}}$}$\; 50\,\mathrm{GeV}$,
and gluon-induced $gg$ and $qg$ processes are expected to
dominate in accessible kinematical regimes. 
%
\begin{figure}[th]
\centerline{\epsfxsize=3.05in\epsfbox{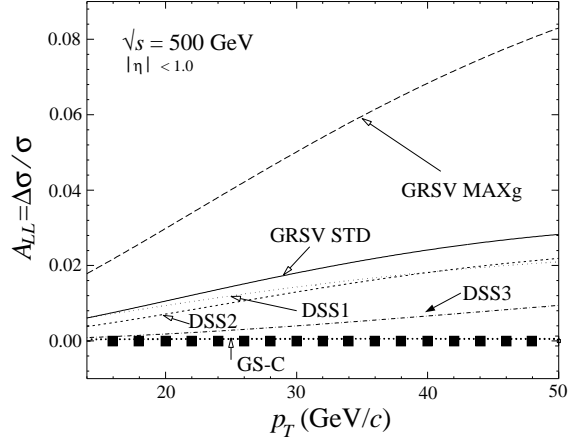}}   
\caption{As in Fig.~\ref{fig:fig3} but now for high-$p_T$
jet production. The ``error bars'', Eq.~(\ref{eq:error}), are for the STAR experiment taking into
account its acceptance. Figure taken from \cite{ref:nlojets}. \label{fig:fig4}}
\vspace*{-0.2cm}
\end{figure}
Due to limitations in the angular
coverage, jet studies will be performed by STAR only. As jet surrogates, PHENIX can 
look for high-$p_T$ leading hadrons, such as pions, whose production proceeds
through the same partonic subprocesses as jet production. Hadrons have the advantage 
that they can be studied also at $\sqrt{S}=200\,\mathrm{GeV}$ and down to lower 
values in $p_T$ than jets as they do not require the observation of clearly 
structured ``clusters'' of particles (jets).
Contrary to jet production, a fragmentation function has to be introduced into the theoretical
framework, cf.\ Eq.~(\ref{eq:eq1}), to take care of final-state collinear singularities.
In case of pion production, the $D_c^\pi$ are, however, fairly well constrained by $e^+e^-$ data. 
It should be also emphasized that in the unpolarized case, the comparison between
NLO theory predictions with jet production data from the Tevatron is extremely successful.
The same is true for first preliminary data on the $p_T$-spectrum for pions at
$\sqrt{S}=200\,\mathrm{GeV}$ from PHENIX \cite{ref:phenixprel}.

\begin{figure}[th]
\centerline{\epsfxsize=3.00in\epsfbox{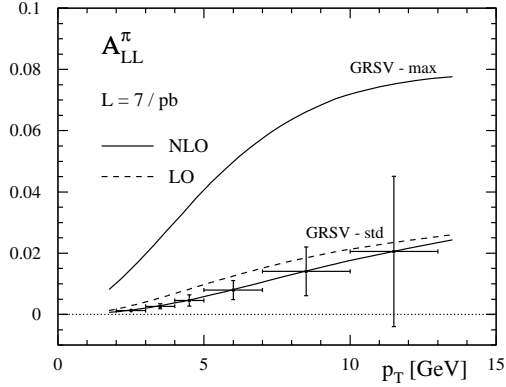}}   
\caption{As in Fig.~\ref{fig:fig3} but now for high-$p_T$
$\pi^0$-production. Note that here the statistical ``error bars'' 
$\delta A_{\mathrm{LL}}$ have been estimated
by assuming only $P=0.4$ and ${\cal{L}}=7\,\mathrm{pb}^{-1}$ 
in Eq.~(\ref{eq:error})
which is a realistic target for the next RHIC run.
Figure taken from \cite{ref:nlohadron2}. \label{fig:fig5a}}
\vspace*{-0.2cm}
\end{figure}

The NLO QCD corrections to polarized jet production are available as a Monte Carlo
code \cite{ref:nlojets}. Apart from a significant reduction of the scale dependence,
they are also mandatory for realistically matching the procedures used in 
experiment in order to group final-state particles into jets.
For single-inclusive high-$p_T$ hadron production the task of computing the 
NLO corrections has been completed only very recently \cite{ref:nlohadron1,ref:nlohadron2}.
Figure~\ref{fig:fig4} shows $A_{\mathrm{LL}}$ for single-inclusive jet production at
the NLO level as a function of the jet $p_T$. A cut in rapidity, $|\eta|\leq 1$, 
has been applied in order to match the acceptance of STAR. The asymmetries turn out to be smaller than
for prompt photon production, but thanks to the much higher statistics one can
again easily distinguish between different spin-dependent gluon densities.
Results for single-inclusive $\pi^0$ production are presented in Fig.~\ref{fig:fig5a}.
Note that here the expected statistical accuracy refers to only a very moderate
integrated luminosity and beam polarization as targeted for the upcoming run of RHIC.
Even under these assumptions a first determination of $\Delta g$ can be achieved.
%

The last process which exhibits a strong sensitivity to $\Delta g$ is heavy flavor production.
Here, the LO gluon-gluon fusion mechanism, $gg\to Q\bar{Q}$, dominates unless
$p_T$ becomes rather large.
Unpolarized calculations have revealed that NLO QCD corrections are mandatory for a meaningful
quantitative analysis. In the polarized case they have been computed  
recently in case of single-inclusive heavy quark production \cite{ref:nlohq}. Again, one observes 
a strongly reduced scale dependence for charm and bottom production at RHIC energies.
It turns out that the major theoretical uncertainty stems from the unknown 
precise values for the heavy quark masses \cite{ref:nlohq}.
Since the heavy quark mass already sets a large scale, one can 
perform calculations for small transverse momenta or even for total cross sections
which, in principle, give access to the gluon density at smaller $x$-values than relevant for 
high-$p_T$ jet, hadron or prompt photon production.

\begin{figure}[th]
\centerline{\epsfxsize=2.60in\epsfbox{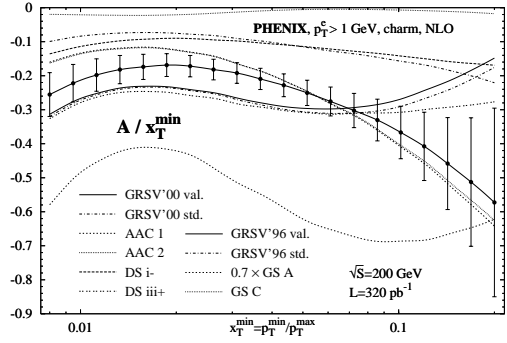}}   
\caption{NLO single-inclusive charm production asymmetry  
(rescaled by $1/x_T^{\mathrm{min}}$) as a function of 
$x_T^{\mathrm{min}}\equiv p_T^{\mathrm{min}}/p_T^{\mathrm{max}}$
for different sets of parton densities. The ``error bars'', Eq.~(\ref{eq:error}), are 
for the PHENIX experiment and include a detection efficiency for 
the channel $c\to eX$. Figure taken from \cite{ref:nlohq}.
\label{fig:fig5}}
\vspace*{-0.2cm}
\end{figure}

Heavy flavors are not observed directly at RHIC but only through their decay 
products. Possible signatures for charm/bottom quarks at PHENIX are 
inclusive-muon or electron tags or $\mu e$-coincidences. The latter provide a much better
$c/b$-separation which is an experimental problem. In addition, lepton detection
at PHENIX is limited to $|y|\leq 0.35$ and $1.2\leq |y|\leq 2.4$ for electrons and
muons, respectively. Since heavy quark decays to leptons
proceed through different channels and have multi-body kinematics, it is
a non-trivial task to relate, e.g., experimentally  observed 
$p_T$-distributions of decay muons to the calculated $p_T$-spectrum of the produced 
heavy quark. One possibility is to model the decay with the help
of standard event generators like PYTHIA \cite{ref:pythia} by computing probabilities
that a heavy quark with a certain $(p_T,y)$ is actually seen within the PHENIX
acceptance for a given decay mode. Figure~\ref{fig:fig5} shows a  
prediction for the charm production
asymmetry $A_{\mathrm{LL}}$ at PHENIX in NLO QCD for the inclusive-electron tag. 
The sensitivity to $\Delta g$ is less pronounced than for the processes discussed
above. It remains to be checked if heavy flavor production at RHIC can be used to 
extend the measurement of $\Delta g$ towards smaller $x$-values.
%

\subsection{Further Information on $\Delta q$ and $\Delta \bar{q}$}
\begin{figure}[th]
\vspace*{-0.3cm}
\centerline{\epsfxsize=0.35\textwidth\epsfbox{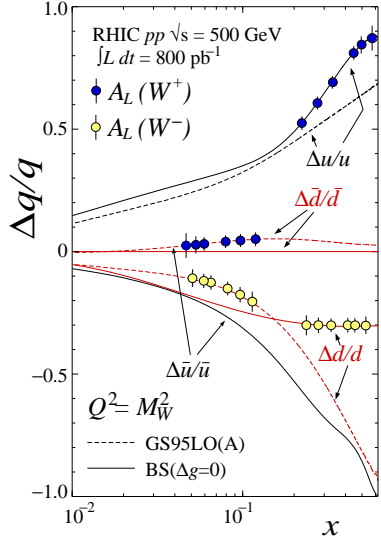}}   
\caption{Expected statistical accuracy for $\Delta q/q$ from
$A_{\mathrm{L}}$ overlayed on two sets of parton densities.
The full [open] circles refer to $A_{\mathrm{L}}(W^+)$
$[A_{\mathrm{L}}(W^-)]$. Figure taken from \cite{ref:rhicreport}. \label{fig:fig6}}
\vspace*{-0.4cm}
\end{figure}

%
Inclusive DIS data only provide information on the sum of quarks and antiquarks
for each flavor, i.e., $\Delta q + \Delta \bar{q}$. At RHIC a separation of
$\Delta u$, $\Delta \bar{u}$, $\Delta d$, and $\Delta \bar{d}$ can be achieved 
by studying $W^\pm$-boson production. 
Exploiting the parity-violating properties of $W^\pm$-bosons,
it is sufficient to measure a single spin asymmetry, $A^W_{\mathrm{L}}$, with only one
of the colliding protons being longitudinally polarized. The idea is to study $A^W_{\mathrm{L}}$
as a function of the rapidity of the $W$, $y_W$, relative to the polarized 
proton \cite{ref:lowboson}.
In LO it is then easy to show \cite{ref:lowboson,ref:rhicreport} 
that for $W^+$-production, $u\bar{d}\to W^+$,
and large and positive (negative) $y_W$, $A^W_{\mathrm{L}}$ is 
sensitive to $\Delta u/u$ ($\Delta \bar{d}/\bar{d}$). Similarly, $W^-$-production probes
$\Delta d/d$ and $\Delta \bar{u}/\bar{u}$. The NLO QCD corrections for $A_{\mathrm{L}}$ 
as well as the factorization scale dependence are small \cite{ref:nlowboson}.
Experimental complications \cite{ref:rhicreport} arise, however, from the fact that 
neither PHENIX nor STAR are hermetic, which considerably complicates the reconstruction of $y_W$.
Therefore it is important to understand $A_{\mathrm{L}}$ on the decay-lepton level.
Here, fully differential NLO cross sections are available as a MC code \cite{ref:pavel}.
The anticipated sensitivity of PHENIX on the flavor decomposed quark and antiquark
densities is illustrated in Fig.~\ref{fig:fig6}.

Semi-inclusive DIS measurements, $ep\to HX$, are another probe to 
separate quark and antiquark densities. HERMES has recently published first
preliminary results \cite{ref:beckmann}. The accessible $x$-range for the
$\Delta q$ and $\Delta \bar{q}$ densities is comparable to that of RHIC,
see Fig.~\ref{fig:fig6}, but at scales $Q\simeq 1-2\,\mathrm{GeV}$ rather than $M_W$.
The combination of both measurements can provide an important test of the QCD 
scale evolution for polarized parton densities and of the possible relevance of higher twist
contributions at low scales.

\subsection{Towards a Global Analysis of Upcoming Data}
%
Having available at some point in the near future data on various different reactions, one
needs to tackle the question of how to set up a ``global QCD analysis'' for spin-dependent
parton densities. The strategy is in principle clear from the unpolarized case:
an ansatz for the densities, Eq.~(\ref{eq:pdfs}), at some initial scale $\mu_0$, given in terms of some
functional form with a set of free parameters, is evolved,
Eq.~(\ref{eq:evol}), to a scale $\mu$ relevant for a
certain data point. A $\chi^2$-value is assigned that represents the quality of the
comparison of the theoretical calculation to the experimental point. The parameters are varied
until eventually a global minimum in $\chi^2$ is reached mutually for all data points.
In practice, this approach is not fully viable since the numerical evaluations of the
cross sections in NLO QCD are usually time-consuming as they require several tedious
integrations. Hence the computing time for a QCD fit easily becomes excessive. 
%

In the unpolarized case, the wealth of DIS data already provides a pretty good 
knowledge of the parton densities, and
reasonable approximations can be made for the most time-consuming processes. 
For instance, one can absorb all NLO corrections into some pre-calculated ``correction
factors'' $K$, and simply multiply them in each step of the fit
to the LO approximation for the cross sections which can be evaluated much faster.
In the polarized case, it is in general not at all clear whether such a strategy will work.
Here, parton densities are known with {\em much} less accuracy so far. It is therefore
not possible to use $K$-factors reliably. In addition, spin-dependent parton densities as
well as partonic cross sections may oscillate, i.e., have zeros, in the kinematical
regions of interest such that predictions at LO and the NLO can show marked differences.
Clearly, in the polarized case the goal {\em must} be to find a way of implementing efficiently,
and without approximations, the {\em exact} NLO expressions for all relevant hadronic
cross sections. A very simple and straightforward method based on ``double Mellin
transformations'' was proposed in \cite{ref:kosower}. Recently, its actual practicability
and usefulness in a global QCD analysis has been demonstrated \cite{ref:global} in a
case study based on fictitious prompt-$\gamma$ data.

\section{Exploring Physics Beyond the Standard Model}
%
Spin observables are also an interesting tool to uncover important new physics.
One idea is to study single spin asymmetries $A_L$ for large-$p_T$ jets. In the 
standard model $A_L$ can be only non-zero for parity-violating interactions,
i.e., QCD-electroweak interference contributions, which are fairly small.
The existence of new parity-violating interactions could lead to sizable
modifications \cite{ref:virey} of $A_L$. Possible candidates 
are new quark-quark contact interactions, characterized by a compositeness scale
$\Lambda$. RHIC is surprisingly sensitive to quark substructure at the 2~TeV
scale, and is competitive with the Tevatron despite the much lower 
c.m.s.\ energy \cite{ref:virey}.
Other candidates for new physics are possible new gauge bosons, e.g.,
a leptophobic $Z'$. Of course, high luminosity and precision as well as
a good knowledge of polarized and unpolarized parton
densities and of the standard model ``background'' are mandatory.
For details, see \cite{ref:virey,ref:rhicreport}.

\section{Summary and Outlook}
%
With first data from RHIC hopefully starting to roll in soon,
we can address many open, long-standing questions in spin physics 
like the longitudinally polarized gluon density. 
With data from many different processes taken at high energies, where perturbative QCD
should be at work, a first global analysis of spin-dependent parton
densities will be possible. For a long time to come RHIC will provide the
best source of information on polarized parton densities,
certainly much improving our knowledge of the spin structure of the nucleon, and, perhaps, 
the next ``spin surprise'' is just round the corner. 
Future projects like the EIC \cite{ref:eic}, which is currently
under scrutiny, would help to further deepen our understanding by probing
aspects of spin physics not accessible in hadron-hadron collisions. The
structure function $g_1$ at small $x$ or the spin content of circularly
polarized photons are just two examples.
  

%
\end{document}